\RequirePackage{iftex}

\ifLuaTeX%
  \RequirePackage[OT1]{fontenc}
\fi%

\documentclass[10pt,conference]{IEEEtran}
\usepackage{amsmath}
\usepackage{hyperref}

\input{bfein-defs}
\NewDocumentCommand{\temet}{ o }{%
  \textsc{TeMet}\IfValueT{#1}{\(_{#1}\)}\xspace
}
\NewDocumentCommand{\fomet}{ o }{%
  \textsc{FoMet}\IfValueT{#1}{\(_{#1}\)}\xspace
}

\newcommand{\dfjAbstractCompilableTestCases}{\num[round-mode=none]{169}\xspace}

\newcommand{\dfjAbstractFailsBuggy}{\num[round-mode=none]{48}\xspace}
\newcommand{\dfjAbstractFailsFixed}{\num[round-mode=none]{64}\xspace}
\newcommand{\dfjAbstractFoundBugs}{\num[round-mode=none]{20}\xspace}

\newcommand{\dfjAbstractNotCompilable}{\num[round-mode=none]{75}\xspace}
\newcommand{\dfjAbstractPassesBoth}{\num[round-mode=none]{57}\xspace}

\newcommand{\dfjAssertionLocationInHelperClosure}{\num[round-mode=none]{246}\xspace}

\newcommand{\dfjAssertionLocationInHelperTotal}{\num[round-mode=none]{277}\xspace}

\newcommand{\dfjAssertionLocationInTestTotalBugs}{\num[round-mode=none]{138}\xspace}
\newcommand{\dfjAssertionLocationInTestTotalTests}{\num[round-mode=none]{244}\xspace}

\newcommand{\dfjRawCompilableTestCases}{\num[round-mode=none]{184}\xspace}

\newcommand{\dfjRawFailsBuggy}{\num[round-mode=none]{55}\xspace}
\newcommand{\dfjRawFailsFixed}{\num[round-mode=none]{84}\xspace}
\newcommand{\dfjRawFoundBugs}{\num[round-mode=none]{33}\xspace}

\newcommand{\dfjRawNotCompilable}{\num[round-mode=none]{60}\xspace}
\newcommand{\dfjRawPassesBoth}{\num[round-mode=none]{45}\xspace}

\newcommand{\dfjTestsInProjectClosure}{\num[round-mode=none]{259}\xspace}

\newcommand{\dfjTotalBugs}{\num[round-mode=none]{434}\xspace}
\newcommand{\dfjTotalTests}{\num[round-mode=none]{1129}\xspace}

\usepackage{url}
\usepackage{booktabs}
\usepackage{balance}

\usepackage{orcidlink}

\title{AsserT5: Test Assertion Generation Using a Fine-Tuned Code Language Model}

\author{%
\IEEEauthorblockN{Severin Primbs \orcidlink{0009-0004-5115-8067}}
\IEEEauthorblockA{\textit{University of Passau} \\
  Passau, Germany \\
}
\and
\IEEEauthorblockN{Benedikt Fein \orcidlink{0000-0002-3798-845X}}
\IEEEauthorblockA{\textit{University of Passau} \\
  Passau, Germany
}
\and
\IEEEauthorblockN{Gordon Fraser \orcidlink{0000-0002-4364-6595}}
\IEEEauthorblockA{\textit{University of Passau} \\
  Passau, Germany
}
}

\begin{document}

\maketitle%

\begin{abstract}
  Writing good software tests can be challenging, therefore approaches
that support developers are desirable. While generating complete tests
automatically is such an approach commonly proposed in research,
developers may already have specific test scenarios in mind and thus
just require help in selecting the most suitable test assertions for
these scenarios. This can be done using deep learning models to
predict assertions for given test code.
Prior research on assertion generation trained these models
specifically for the task, raising the question how much the use of
larger models pre-trained on code that have emerged since then can
improve their performance. In particular, while abstracting
identifiers has been shown to improve specifically trained models, it
remains unclear whether this also generalises to models pre-trained on
non-abstracted code. Finally, even though prior work demonstrated high
accuracy it remains unclear how this translates into the effectiveness
of the assertions at their intended application -- finding faults.
To shed light on these open questions, in this paper we propose
\assertfive, a new model based on the pre-trained \codetfive model,
and use this to empirically study assertion generation.
We find that the abstraction and the inclusion of the focal method are
useful also for a fine-tuned pre-trained model, resulting in test
assertions that match the ground truth assertions precisely in up to
59.5\,\% of cases, more than twice as precise as prior models.
However, evaluation on real bugs from the \defectsforj dataset shows
that out of \dfjAssertionLocationInTestTotalBugs{} bugs detectable
with assertions in real-world projects, \assertfive was only able to
suggest fault-finding assertions for \dfjRawFoundBugs, indicating the
need for further improvements.

\end{abstract}


\section{Introduction}

Developing reliable and high-quality software is a time-consuming and
resource-intensive process~\cite{Yang2006}. An important part of
creating such high-quality software is extensive testing of the
implemented functionality since the validation of program source code
plays a crucial role in identifying and eliminating potential errors
at an early stage~\cite{Beck2000}.

Since testing software is a time-consuming task, tools such as
\evosuite~\cite{Fraser2011} or
\properNoun{Randoop}~\cite{Pacheco2007}, and recently also tools based
on large language models~(\eg~\cite{Nie2023,Schaefer2024,He2024}), aim
to automatically generate test methods. However, sometimes developers
already have specific test scenarios with the setup of specific test
states in mind. A challenging step then is to generate test assertions
that check whether the developers' assumptions about the program state
hold.
We therefore aim to automatically generate new or additional
assertions within pre-existing test methods.

\begin{figure}[t]
  \begin{subfigure}{0.45\linewidth}
    \centering%
    \begin{lstlisting}[aboveskip=0pt,belowskip=0pt]
  char last(String s) {
    return s[s.length-1];
  }
    \end{lstlisting}
    \caption{%
      Focal method.
    }\label{fig:intro-example-focal-method}
  \end{subfigure}
  \hfill%
  \begin{subfigure}{0.45\linewidth}
    \centering%
    \begin{lstlisting}[aboveskip=0pt,belowskip=0pt]
@Test void testLast() {
  char res = last("abc");
  assertEquals(res, 'c');
}
    \end{lstlisting}
    \caption{%
      Its unit test method.
    }\label{fig:intro-example-test}
  \end{subfigure}
  \hfill%
  \begin{subfigure}{1.0\linewidth}
    \begin{lstlisting}[breakindent=0pt,breakautoindent=false,belowskip=0pt]
char last(String s) { return s[s.length -1]; } <SEP> @Test void testLast() { int res = last("abc"); <ASSERTION> }
    \end{lstlisting}
    \caption{As input to an assertion generation model.}\label{fig:intro-example-input}
  \end{subfigure}
  \caption{%
    Example Java method and its unit test when given to a deep-learning-based assertion generation model that is trained to replace the \texttt{<ASSERTION>} placeholder.
  }\label{fig:example-assertion-generation}
\end{figure}

One possible approach to produce meaningful test assertions is to use
artificial intelligence to predict assertions from the source code. As
shown in \cref{fig:example-assertion-generation}, such a predictive
model receives the text of the test
method~(\cref{fig:intro-example-test}) and the method under test~(also
called the focal method, \cref{fig:intro-example-focal-method}) as its
input and is tasked with generating another text sequence containing
the assertion statement to fill in the placeholder. The model
therefore has to learn to reason about the semantics of the code
purely from its textual representation.
This approach has been employed successfully for example with the
\atlas~\cite{Watson2020} and \toga~\cite{Dinella2022} assertion
generation models.

Since the models work on the textual representation, they need to
learn to reason about a large vocabulary of constants and identifiers.
To support this in their model, \atlas integrated an abstraction
process that replaces such identifiers with abstract tokens to
decrease the vocabulary size. As an alternative, the use of larger
pre-trained natural language models that have been fine-tuned
twice~(on source code, then for assertions) has also been investigated
to alleviate the vocabulary problem~\cite{Tufano2022}.
However, since the introduction of these approaches, larger models
pre-trained on code have been released. Because such models are
trained on raw code without any abstractions, it is unclear whether
pre-trained models would need to be fine-tuned with abstracted or raw
tokens.
Furthermore, while prior results show that the focal method context
improves the prediction performance~\cite{Tufano2022}, it remains
unclear if other pre-existing assertions in the test method can
provide a sufficient alternative source of information.
Finally, approaches have also mainly been evaluated according to
common machine learning
metrics~\cite{Watson2020,Wang2024,Mastropaolo2021}, rather than by
demonstrating whether the assertions are effective at finding faults;
this has so far only been evaluated using artificially generated
tests~(\eg~using \evosuite)~\cite{Tufano2022,Dinella2022}, rather than
developer written tests.

To shed light on these open questions, we introduce \assertfive, a new
model for assertion generation that combines the features of the
previous approaches while alleviating the individual limitations by
using a larger language model that was pre-trained on source code and
then fine-tuned for the task of assertion generation.
The pre-trained code model does not need to learn the code structure
from scratch %
but instead can learn how
different abstract identifiers relate to each other during the
fine-tuning.
This base model requires only a single fine-tuning step on assertions,
thus avoiding catastrophic forgetting~\cite{Bower1989}, and
it already is designed to generate sequences of source code, allowing
us to generate both regular assertions statements and special ones
expecting an exception without requiring a grammar like prior
work~\cite{Dinella2022}.

In detail, the contributions of this work are:
\begin{itemize}
  \item We present the \assertfive assertion generator based on a
    fine-tuned \codetfive model and compare it against current
    state-of-the-art approaches.
  \item We show that pre-existing assertions in the test case improve
    the model performance in cases where the focal method cannot be
    determined.
  \item We evaluate the bug detecting capabilities by integrating
    generated assertions into developer-written tests.
\end{itemize}

\section{Background}

\subsection{Deep-Learning-Based Assertion Generation}

As demonstrated by the example in
\cref{fig:example-assertion-generation}, deep learning models used to
generate assertions are usually text-based. They receive the source
code of the test method and optionally the code of the focal method to
generate the source code representing the assertion.
This task is therefore suitable for sequence-to-sequence
models~\cite{Britz2017}. Initial approaches like
\atlas~\cite{Watson2020} trained a dedicated recurrent neural network
to generate assertions. However, since code contains many different
project-specific identifiers and constants, the vocabulary that has to
be managed by the model is quite large unless vocabulary reducing
techniques such as using abstract tokens instead of concrete constants
and identifiers are introduced~\cite{Watson2020}.

Therefore, newer approaches tend to use larger
Transformer-based~\cite{Vaswani2017} architectures that are better
suited to efficiently handle such large vocabularies. By using a
generic Transformer base-model that has been pre-trained either on
natural language or source code, it only needs to be fine-tuned when
used as part the final assertion generation model. This shortens the
required training time and allows using larger, more expressive,
models.
Still, there are differences in how the underlying architecture is
integrated into the assertion generation model.
For example, the ability to fine-tune models allows for training a
\bart Transformer~\cite{Lewis2020} on large English natural language
dataset, then tuning it on source code, before finally tuning it again
to generate the assertion statements as required for the actual
task~\cite{Tufano2022}.
Other Transformer architectures like \bert~\cite{Devlin2019} are not
designed to generate sequences, but instead are used to classify the
input sequences. Combined with the insight that many assertion
statements follow a fairly rigid structure, such models are used as
part of the \toga~\cite{Dinella2022} approach. Instead of letting the
model freely generate source code, the approach generates a set of
valid assertion templates according to the allowed structures that can
be filled in with elements from the surrounding test method. The \bert
classifier therefore only needs to suggest the most suitable
template~\cite{Dinella2022}.

\subsection{T5 Transformer Models}

The Text-To-Text Transfer Transformer~(\tfive)~\cite{Raffel2019}
follows the encoder-decoder Transformer architecture and therefore is
designed for encoding input sequences into internal vector
representations which then are decoded back into output sequences.
It has been successfully used to generate sequences of source code
after pre-training the model on mixed sets of natural language and
Java source code before fine-tuning exclusively on
code~\cite{Mastropaolo2021}.

To create a model specifically designed to facilitate code-related
tasks, \codetfive~\cite{Wang2021}, a different set of pre-training
tasks like restoring the names of masked identifiers are used.
\codetfivelarge~\cite{Le2022} builds upon the same architecture but
introduces improved pre-training strategies, such as enhanced learning
objectives, expanded model sizes, and better datasets.

\section{AsserT5}

Our proposed model, \assertfive, is built on top of a pre-trained
Transformer architecture. The model receives a unit test together with
its focal method as a single input sequence and is tasked to generate
an assertion statement.
Due to the pre-trained base-model, we focus on fine-tuning the
underlying Transformer on a dataset of pairs of tested methods and
their corresponding test cases. We use a pre-existing dataset of such
pairs as basis, but also apply additional filtering and preprocessing
steps to adapt the data points to our requirements.

\subsection{Base Dataset: Methods2test}\label{sec:methods2test}

\Methodstotest~\cite{Tufano2022a} is a large, supervised comparison
dataset that assigns Java \junit tests to their associated focal
methods. The dataset aims to find a reliable mapping that ensures that
the focal method belongs to the associated test. Originally intended
to automatically generate test cases~\cite{Tufano2020}, \methodstotest
fills the gap of missing datasets with real test cases. It contains
780k data elements that contain \junit tests and their focal method
from 9.4k Java open-source non-fork projects from \github that the
maintainer updated in the last five years.

While preprocessing a project into the structure desired by
\methodstotest, each project is parsed to identify the test and focal
methods using the following heuristics:
The class containing the focal method must have the name of the
corresponding test class without the \enquote{test} prefix or suffix
and be part of the same Java package. The search for the focal method
continues only in the found focal class.
Removing a \enquote{test} prefix or suffix from the test method names
and finding this modified method name in the focal class yields the
focal method.
If this does not work, \methodstotest extracts all methods called in
the test case and the focal class and forms the intersection of these
sets. If this intersection contains exactly one element, this is the
focal method. Otherwise, no focal method for the test method is
detectable.
\Methodstotest discards the data point if no unique focal method
exists after these strict heuristics.

Other datasets were generated using simpler focal method detection
heuristics like choosing the last method call before the
assertion~\cite{Watson2020}. However, this assumption has been found
to not reflect the developer intention in many cases~\cite{He2024}.
The more complex set of heuristics of \methodstotest therefore ensures
better matching test and focal method pairs.

\subsection{Training and Evaluation Dataset}\label{sec:preprocessing}

We used the \methodstotest data as basis for our extended dataset.
Since the dataset is missing the Javadoc documentation for the focal
method which we require later to compare our model with
\toga~\cite{Dinella2022}, we cloned all still available source
repositories again~(\exnum{9184} cloneable, \exnum{226} not) to
extract the original test and focal method with the additional
context.

\subsubsection{Data Filtering}\label{sec:data-filtering}

We discarded data points where either the test or focal method is no
longer available to obtain \exnum{562836} out of the original
\exnum{780944} data points from which we removed further \exnum{4402}
samples where the focal method is a constructor rather than an actual
method.
To later understand how the amount of context affects the performance
of the model, we generate three subsets from this filtered dataset.
The first subset allows exactly one assertion per test
case~(\exnum{149893} tests), the second one additionally adds tests
with up to five assertions~(\exnum{248831} tests), and for the third
one we allowed up to ten assertions~(\exnum{269490} tests).
To mask the assertions in the test method and to extract the focal
method, the classes for both need to be parseable into an abstract
syntax tree~(\AST). This requirement removed \exnum{6701} data points,
which the parser library we used could not process, from the dataset.
Finally, we removed \exnum{322} items with test cases longer than
\exnum{10000} characters from the dataset to improve performance
during preprocessing.

\begin{table}[t]
  \caption{%
    Assertion types we considered for our datasets with exactly one, up to five, or up to ten assertions in each test case.
  }\label{tab:considered-assertion-types}
  \centering%
  \footnotesize%
  \begin{tabular}{lcrrr}
    \toprule
                   &              & \multicolumn{3}{c}{Frequency in Dataset}\\
    Assertion Type & \#Parameters & \multicolumn{1}{c}{1} & \multicolumn{1}{c}{\(\leq 5\)} & \multicolumn{1}{c}{\(\leq 10\)}\\
    \midrule
    \texttt{assertEquals}              &   2 & \perc{58.34} & \perc{58.49} & \perc{59.33}\\
    \texttt{assertNotEquals}           &   2 &  \perc{0.39} &  \perc{0.56} &  \perc{0.63}\\
    \texttt{assertTrue}                &   1 & \perc{15.36} & \perc{17.74} & \perc{17.85}\\
    \texttt{assertFalse}               &   1 &  \perc{7.12} &  \perc{7.87} &  \perc{8.23}\\
    \texttt{assertNull}                &   1 &  \perc{5.24} &  \perc{4.19} &  \perc{3.92}\\
    \texttt{assertNotNull}             &   1 &  \perc{5.73} &  \perc{7.04} &  \perc{6.69}\\
    \texttt{assertThrows}              &   2 &  \perc{2.42} &  \perc{1.10} &  \perc{0.84}\\
    \texttt{try-catch} + \texttt{fail} & --- &  \perc{5.40} &  \perc{3.01} &  \perc{2.50}\\
    \bottomrule
  \end{tabular}
\end{table}

The main filtering step only accepts data points that have assertions
in the format shown in \cref{tab:considered-assertion-types}.
Like \toga~\cite{Dinella2022}, we only allow commonly used \junit
assertion types and also add \texttt{assertNotEquals} to consistently
allow the positive and negative counterparts and \texttt{assertThrows}
as built-in alternative to the \texttt{try-catch} assertion. This type
verifies that an exception is thrown by the method under test and
optionally makes additional assertions on the caught exception.
Most \junit assertions also allow an additional optional parameter
that contains an error message shown to the developer in case of
assertion failures. This parameter does not influence the values the
assertion is applied to. Therefore, we only consider assertions
without this parameter to later ease the automatic evaluation if two
assertions are functionally equivalent.

Each of the three datasets with different numbers of assertions is
independently split into training, validation, and test sets following
an 80:10:10 ratio. We ensure that focal methods with multiple test
cases do not appear in both the training and test splits to avoid
leaking information between splits.
Since in the latter two subsets each test case can have multiple
assertions matching our requirements, we added it to the final dataset
once per valid assertion and masked only one specific assertion per
data point. To avoid data leakage, we again ensured all variants
appeared in only one of the train/eval/test splits.
This resulted in \exnum{140897}, \exnum{411132}, and \exnum{555883}
usable data points in the subsets with 1, 5, and 10 assertions,
respectively.

\subsubsection{Data Preprocessing}\label{sec:data-preprocessing}

This filtered dataset needs to be transformed into a customised format
specific to the assertion generation model. That entails concatenating
all source code tokens from the test and focal methods.

We constructed three dataset variants for the evaluation.
The first preprocessing variant comprises raw text tokens exclusively
(\ie~tokenising the code without further changes to the actual
tokens), forming a raw dataset by concatenating the test and focal
method.
The second option uses the same tokenisation steps but only considers
the test method code.
The third variant uses the abstract tokens of the test and focal
methods. This methodology employed in \atlas aims to abstract the
dataset and entails converting each method name, identifier, or
literal type into an abstract token with a corresponding type and
number. We also included the tokens of the test and focal class to
have more reasonable abstract tokens in the vocabulary. The
potentially important syntactic positions of the abstract tokens
relative to each other remain as in the original code.
The abstraction process augments the model’s capacity to discern and
generalise patterns and features of the data~\cite{Tufano2019,
Tufano2018, Tufano2019a, Watson2020}\@. \atlas has shown that
abstraction improved evaluation scores compared to the raw
variant~\cite{Watson2020}, as the abstraction process reduced the
vocabulary size and therefore the number of model parameters as well
as the training duration.

\begin{figure}[t]
  \begin{lstlisting}[basicstyle=\ttfamily\fontsize{6.7}{7}\selectfont,aboveskip=0pt,belowskip=0pt]
TEST_METHOD: @ Test void METHOD_2 ( ) { char IDENT_1 = METHOD_0 ( STRING_0 ) ; <ASSERTION> }
FOCAL_METHOD: char METHOD_0 ( String IDENT_0 ) { return IDENT_0 [ IDENT_0 . METHOD_1 - INT_0 ] ; }
ASSERTION: ASSERT_0 ( IDENT_0 , CHAR_0 )
  \end{lstlisting}
  \caption{%
    Abstraction of the token sequence from \cref{fig:example-assertion-generation}.
  }\label{fig:tokenised-abstract-method}
\end{figure}

We use the test method \texttt{testLast} and its focal method from
\cref{fig:example-assertion-generation} to demonstrate how the
abstraction process replaces all semantic identifiers, method names,
and literal values during input generation for model training and
inference.
As \cref{fig:tokenised-abstract-method} shows, the starting points of
the test and focal methods are marked before concatenating test and
focal methods.
Then identifiers and constant values within the code are transformed
into abstract tokens. We encode string literals that contain
whitespace into exactly one abstract token rather than splitting them
into many small subtokens.
For all replacements, the original values of the replaced tokens are
saved in a dictionary specific to this individual test and focal
method pair.
For example, the abstract token \texttt{INT\_X} corresponds to the
number 5 in the shown test case, but may refer to another integer
constant in another input. This forces the model to learn from a more
general structure of the inputs rather than relying on specific
constants or names.
When the model generates a sequence of tokens during training or
inference, it then has to only choose between a comparatively small
set of possible tokens.
While the input-specific stored dictionary of identifiers is not used
during training since both input and output only use the abstract
form, it is used during inference to map the abstract tokens back to
proper values, \ie~usable source code.

In the model input, the assertion is removed and instead masked by
\texttt{<ASSERTION>} during training. The abstracted assertion is not
part of the model input but still presented in
\cref{fig:tokenised-abstract-method} to show the ground truth label we
expect the model to predict. To obtain the same structure during
inference, the special token is added to the location in which we want
the model to generate the assertion statement.
If the final model input sequence is longer than the maximum supported
length of \(n = 386\) tokens, we truncate the sequence so that only
the first \(n\) tokens of the sequence are passed to the model and
discard the remainder, but always retain the assertion placeholder.

\subsection{Model}\label{sec:model}

\assertfive is based on a fine-tuned \codetfive~\cite{Wang2021} model.
The underlying \tfive model~\cite{Raffel2019} can capture long
dependencies between tokens which allows the generated assertions to
reference back to variable names appearing in the test code.
Choosing \codetfive for fine-tuning was a strategic decision rooted in
several practical key factors:
(1)~\emph{Pre-training}:~\codetfive is specifically designed and
pre-trained on a diverse set of code-related tasks, giving a strong
foundation for understanding programming languages, syntax, and
semantics. As a sequence-to-sequence model it therefore is designed to
generate code.
(2)~\emph{Existing model basis}:~\huggingface provides a model basis
for creating text
sequences\footurl{https://huggingface.co/docs/transformers/v4.37.2/en/model_doc/t5}[2025-01-20],
which is ideal for generating test assertions.
(3)~\emph{Realistically trainable}:~The model with approximately 770
million parameters is trainable on a single \properNoun{Nvidia}
A100-80GB \gpu, which allows us to reasonably train and compare
multiple model variations.
(4)~\emph{Scalability}:~\codetfivelarge is smaller than comparable
models while at the same time outperforming larger
models~\cite{Le2022}. Due to its comparatively small size it is also
fast enough during inference when generating many test assertions.

We used the pre-trained \codetfivelarge
model\footurl{https://huggingface.co/Salesforce/codet5-large}[2025-01-20]~\cite{Le2022}
which can generate text sequences~(in this case, assertions) based on
the input tokens. We used the architecture of
\codetfivelarge~\cite{Le2022} without changes and allowed each
parameter to be trainable, \ie~we did not freeze any layers during
fine-tuning.
As optimiser, we used AdamW~\cite{Loshchilov2017} with a learning rate
of \exnum{2e-5}. The rest of the hyperparameters stayed at the default
values. We scheduled our training process with a linear scheduler and
used no warm-up steps.
We trained our model for ten epochs and used a batch size of 38 to
fill the available \qty[round-mode=none]{80}{\gibi\byte}
\initialism{gpu} memory.
To determine the input and output sequence lengths, we observed that
the concatenated test and focal method sequences are usually longer
than the assertion statement. Therefore, we adapted the sequence
lengths accordingly to allow up to \exnum{386} input tokens and up to
\exnum{64} output tokens.
While increasing the input and output token lengths would allow for
larger test methods and assertions without having to truncate them, it
would also increase the training duration since longer sequences need
additional \initialism{gpu} space and therefore necessitate the use of
a smaller batch size.

Following the two different preprocessing variants once using concrete
tokens and once using abstracted variants, we trained two different
model variants to investigate whether the improvement through the
abstraction process observed for \atlas~\cite{Watson2020} also occurs
for our Transformer-based model.
Prior research suggests that the byte-pair-encoding employed by
\properNoun{T5} to tokenise the code can effectively avoid
out-of-vocabulary~(\textsc{oov}) situations in code completion
scenarios~\cite{Karampatsis2020}.
However, even if not required to mitigate the \textsc{oov} problem, we
consider both the model variant with and without abstracted tokens in
our evaluation since the assertion statements follow a fairly strict
structure. Therefore, the abstracted tokens might still result in an
improvement in the prediction performance since the model can focus on
learning useful structures during the fine-tuning.

\section{Evaluation}

We aim to answer the following research questions:

\begin{description}

\item[RQ1] How does the context of the focal method and other
  assertions in the test affect the model performance?

\item[RQ2] How does the performance of \assertfive compare to existing
  approaches?

\item[RQ3] Does \assertfive generate fault-detecting assertions for
  developer-written tests?

\end{description}

\subsection{RQ1: Relevance of Context}\label{sec:rq-context}

To successfully create useful assertions as a developer, not only the
test method but also an understanding of the method under test is
important. In this research question we answer if this additional
context also helps the \assertfive model during the automatic
assertion generation.
Since it is often difficult to determine the corresponding test method
in the practical application of the models (for example, in an \ide
plugin), it is of practical relevance to what extent the model still
works if the context of the focal method is unavailable.
Since the model has to rely exclusively on the test method code as
context in such cases, we also evaluate whether other pre-existing
assertions in the test case can replace the missing context
sufficiently.

\subsubsection{Experimental Setup}\label{sec:rq-context-setup}

As described in \cref{sec:data-filtering}, we created three datasets
with up to one, five, or ten assertions in each test method,
respectively.
For all three of those subsets, we use the variant using the raw
rather than the abstract tokens~(see \cref{sec:data-preprocessing}) to
retain the original context of the user-defined identifiers.
For each of the three datasets, we trained two model variants where
the first one receives only the tokens of the test method as input,
while the second one receives the tokens of both test and focal
methods.
This allows us to look at two different types of context: the dataset
choice demonstrates if the model can learn from possibly similar
assertion examples in the test method and the model input variant
highlights the importance of the context available from the method
under test.
In the further course of this section, we call the model variant that
only receives the test method \enquote{\temet{}} and the one that also
receives the focal method \enquote{\fomet{}}. When referring to a
specific model instance, an index indicates the used training dataset,
\eg~\temet[5] was trained on the dataset containing test methods with
up to five assertions each.

To compare the performance of the model variants, we use metrics
frequently used in the machine learning context.
The \topk{k} accuracy compares how often the model can predict
assertions fully matching the original. Looking at the precision,
recall, and F1 scores of the prediction of the assertion type~(see
\cref{tab:considered-assertion-types}) highlights if the model
understood enough about the code to at least suggest the correct
method.
Evaluating the \bleu scores highlights how close the generated
assertions are to the ground truth even when not achieving perfect
matches.
Specific to code models, we evaluate how often the model generates
syntactically correct Java code.

\subsubsection{Threats to Validity}

A threat to construct validity may arise from the maximum input
sequence length of 386 in both models. Since the input sequences for
\fomet are longer than for \temet, sometimes parts of the focal method
had to be truncated. This may inhibit the ability for \fomet to
capture all relevant semantics of the input.
In the dataset with one assertion, the data for \temet had an average
length of 95.5 (median: 65), and \perc{2.1} of the input data got
truncated. Equivalently, inputs for \fomet had an average length of
230.5 (median: 162) which resulted in a truncation for \perc{14.2} of
the inputs.
Using a longer input token sequence length was infeasible regarding
training duration~(see \cref{sec:model}). Therefore, \fomet might
underperform in our evaluation compared to an otherwise identical
model that allows longer input sequences.

\subsubsection{Results}

\begin{table}[t]
  \caption{%
    Influence of adding the focal method in the input~(\textsc{FoMet}) compared to the model variant only receiving the test method as context~(\textsc{TeMet}).
    All values in \%.
  }\label{tab:comparison-focal-method}
  \centering%
  \scriptsize%
  \setlength{\tabcolsep}{5pt}
  \begin{tabular}{lcccccc}
    \toprule
    & \multicolumn{2}{c}{\textbf{One Assertion}}&\multicolumn{2}{c}{\textbf{Five Assertions}} &\multicolumn{2}{c}{\textbf{Ten Assertions}} \\
    & \textsc{TeMet} & \textsc{FoMet} & \textsc{TeMet} & \textsc{FoMet} & \textsc{TeMet} & \textsc{FoMet} \\
    \midrule
    \textbf{Accuracy}        &                &                &                &                &                &                \\
    \ \ \topk{1}             & 37.23          & 43.95          & 45.49          & 49.38          & 47.81          & 51.21          \\
    \ \ \topk{5}             & 47.02          & 55.29          & 58.52          & 62.80          & 61.25          & 64.80          \\
    \ \ \topk{10}            & 49.41          & 57.77          & 61.30          & 65.86          & 64.26          & 67.78          \\
    \textbf{\bleu}           & 78.57          & 82.54          & 84.82          & 86.54          & 86.40          & 87.73          \\
    \textbf{Assertion Type} &                &                &                &                &                &                \\
    \ \ Precision            & 75.47          & 82.99          & 81.91          & 82.05          & 83.48          & 85.06          \\
    \ \ Recall               & 70.64          & 78.65          & 75.21          & 79.29          & 78.76          & 81.33          \\
    \ \ F1                   & 72.87          & 80.55          & 78.10          & 80.58          & 80.95          & 83.09          \\
    \textbf{Syntactic Corr.} & 99.34          & 99.23          & 99.47          & 99.53          & 99.71          & 99.63          \\
    \bottomrule
  \end{tabular}
\end{table}

\cref{tab:comparison-focal-method} shows the performance of \temet and
\fomet. On the smallest dataset, \temet[1] had a \topk{1} accuracy of
\perc{37.23} but was clearly outperformed by \fomet[1] which achieved
an exact match for \perc{43.95} of the samples.
\temet benefitted more from an increased number of assertions to
improve the score by \perc{10.58}-points to \perc{47.81}, while \fomet
only achieved an improvement by \perc{7.26}-points to \perc{51.21}.

The score assimilation pattern repeats for the \bleu scores where
\temet[1] achieved a score of \perc{78.57} when evaluating the dataset
with one assertion~(\fomet[1]: \perc{82.54}).
The precision, recall, and F1 scores show that \fomet outperformed
\temet when considering only the assertion type of the prediction.
The additional context has no relevant influence on the models’
ability to generate syntactically correct assertions with both models
nearly always producing parseable Java code.

The \temet[5] variant achieves better accuracy than \fomet[1], which
suggests that the context from pre-existing assertions in the test can
compensate for the lack of focal method context. While further
increasing the number of assertions in the context improves the
results for \temet[10], the worse results compared to \fomet[5] show
that at this point the focal method provides more relevant information
than the additional assertions.

The values for the datasets with multiple assertions might be closer
together because the tokeniser has to truncate more parts of the focal
method due to the longer input.
An alternative explanation for the values moving closer together could
be that there are more assertions that can be used by the models to
create useful assertions similar to the already existing ones without
having to rely on the context of the focal method.
This importance of assertions serving as example is supported by the
close \bleu scores in case of ten assertions since \temet predicts
assertions closely matching the original even without the focal method
context.

Overall, the results show that additional assertions already present
in the test do not \enquote{distract} the model from predicting
additional different assertions but on the contrary allow the model to
improve from the additional context.
In practice, it might not be always possible to automatically
determine a relevant focal method using
heuristics~\cite{He2024,Tufano2022a}. Our results confirm previous
results that using the method under test as additional input is
beneficial~\cite{Tufano2022}, but also show that the model can still
produce useful suggestions without.
Additionally, our results show that other assertions in the test case
can provide sufficient context for the model to offset the negative
impact of a missing focal method context.

\summary{1}{\assertfive performed better when we added the focal
method to the input sequence. Pre-existing assertions in the test case
can provide a similar improvement to the model performance as the
focal method context. Combining both sources of context results in the
overall best results.}

\subsection{RQ2: Comparison to Existing
Approaches}\label{sec:rq-comparison}

In this research question, we explore how \assertfive compares to
state-of-the-art dedicated assertion generation models
\atlas~\cite{Watson2020}, \toga~\cite{Dinella2022}, pre-trained
\bart~\cite{Tufano2022}, and the general-purpose large language
model~(\llm) \gptfourmini.

\subsubsection{Experimental Setup}

\begin{figure*}[t]
  \scriptsize
  \setlength{\tabcolsep}{3pt}

  \begin{tabular}{lp{17.3cm}}
    \(S\)
    & 
    You will receive two code snippets that are written in the Java programming language.
    The first code snippet contains a test method, and the second code snippet is the focal method that is exercised by the test method.
    The test method snippet contains a masked \enquote{\(<\)ASSERTION\(>\)} part.
    Please suggest 10 different and suitable assertions for this masked statement, ranked by their suitability.
    Only return Java code!
    Only use the JUnit assertion methods \enquote{assertTrue}, \enquote{assertFalse}, \enquote{assertEquals}, \enquote{assertNotEquals}, \enquote{assertNull}, \enquote{assertNotNull}, \enquote{assertThrows}.
    Alternatively, assert expected exceptions using a try-catch and the \enquote{fail} method.
    Add an empty line between assertions.\\

    \(P\)
    &
    Focal method: \verb|'''{{ focal_method_code }}'''| Test method: \verb|'''{{ test_method_code }}'''|\\
  \end{tabular}
  \caption{System message \(S\) and prompt template \(P\) for \chatgpt with placeholders for focal and test methods.}\label{fig:gpt-prompt}
\end{figure*}

Some of the models we compare against required adaptions to the model
architecture or the data preprocessing to be usable as part of the
experiment.

\paragraph{Models}

The \atlas model~\cite{Watson2020} is accompanied by a replication
package %
which contains all the relevant training settings and scripts to start
the model training.
However, the code for training the sequence-to-sequence model is
missing in its repository. Deducing the originally used implementation
from documentation we integrated the \seqtoseq
library~\cite{Britz2017}.
Since the hyperparameters used to train \atlas are not specified in
the paper, we used the values available in the replication package
under the assumption that they represent the final optimal ones.
We increased the model training duration from \num{300000} to
\num{500000} steps since we noticed that the model had not learned
sufficiently in the shorter span but otherwise kept the default model
parameters of \atlas.
Furthermore, since \atlas also supports the raw and abstract variants,
we trained models on both datasets~(see
\cref{sec:data-preprocessing}).

For the double-pre-trained \bart transformer
model~\cite{Tufano2022}~(in the following called \doprebart) no
replication package was available.
We therefore asked the authors %
for guidance on
how to replicate the architecture and followed their recommendations.
The
\bart{}\footurl{https://huggingface.co/facebook/bart-large}[2025-01-20]
model~\cite{Lewis2020}, which had been pre-trained in English
language, was used as the base model. It was then fine-tuned on source
code using the \codesearchnet~\cite{Husain2019} dataset for
Java\footurl{https://huggingface.co/datasets/code_search_net/}[2025-01-20].
The pre-training finished after ten epochs, and we employed this
model checkpoint as a starting point for the second fine-tuning on our
assertions dataset in the raw variant to fine-tune for an additional
ten epochs.

We modified the process of \toga~\cite{Dinella2022} to \togaast.
The fine-tuned \bert model~\cite{Devlin2019} used by \toga to predict
the \texttt{try-catch} type of assertions remains unchanged.
For regular assertions, \toga uses another classifier to select the
best assertion from the given variants. However, this structure does
not allow for a \topk{k} selection of the assertions. To be able to
include \toga into our model comparison, we therefore replaced this
\bert classifier by a \bart~\cite{Lewis2020} sequence generation
model, \ie, \togaast no longer uses the assertion templates to generate
the assertion but retains the two-step decision which kind of
assertion should be generated.
We fine-tuned the two dedicated \bert and \bart models for ten epochs
and used the model with the best validation scores.

For the comparison with \chatgpt~\cite{Brown2020}, we created the
prompt template shown in \cref{fig:gpt-prompt}. The prompt uses a
short introduction explaining the task and expected response format to
the \llm. The zero-shot approach not relying on giving examples to the
\llm has been shown to perform better in similar
scenarios~\cite{Wang2024}.
Then, we tried to extract the assertions for each question discarding
the ones not following a sufficiently structured format. We required at
least ten assertion suggestions, to allow for a meaningful \topk{k}
analysis and removed all responses with fewer suggestions. We also
checked whether the returned code snippets corresponded to one of the
allowed assertion methods or the \texttt{try-catch} structure.
This resulted in \exnum{26963} usable responses.
We used the \properNoun{OpenAI} \api to query the
\textit{gpt-4o-mini-2024-07-18} model between 3 and 6 September~2024
using a temperature of 1.

\paragraph{Data Preprocessing}

To ensure a consistent \AST and thereby remove a possible confounding
factor~\cite{Utkin2022}, we reimplemented the transformation from
source code into the model-specific input format in the same tool we
also use for \assertfive~(see \cref{sec:data-preprocessing}).
We only use the dataset that allows up to ten assertions per test case
for this evaluation.

For \atlas, we used the raw and abstract variants with the
concatenated test and focal method.
Since \doprebart does not support an abstracted variant, we only used
the raw dataset also with concatenated test and focal method.
For the comparison with \chatgpt~\cite{Brown2020}, we exported the
test method, the focal method, and the expected assertion and created
prompts following the template shown in \cref{fig:gpt-prompt} which we
then sent to the \properNoun{OpenAI} \api.

For the \togaast preprocessing, we adapted the \toga steps slightly.
Like the original, we split the dataset into \texttt{try-catch}
assertions and regular assertion methods.
We construct inputs for regular assertions methods by concatenating
test method, focal method, and if available the method-level
documentation of the focal method. This sequence is truncated from the
end if it does not fit into the model input.
For the \texttt{try-catch} assertions, we retained the original
mechanism of concatenating test and focal method with a special
separator token in between and truncating both equally if necessary.
Finally, the two models to generate \texttt{try-catch} assertions and
to generate the regular assertions were trained separately on the
relevant subsets of the overall dataset.

\paragraph{Evaluation Metrics}

To compare the performance of the model variants, we use metrics
frequently used in the machine learning context.
The \topk{k} accuracy compares how often the model can predict
assertions fully matching the original. Looking at the precision,
recall, and F1 scores of the prediction of the assertion type~(see
\cref{tab:considered-assertion-types}) highlights if the model
understood enough about the code to at least suggest the correct
method.
Evaluating the \bleu scores highlights how close the generated
assertions are to the ground truth even when not achieving perfect
matches.
Specific to code models, we evaluate how often the model generates
syntactically correct Java code by checking if the generated assertion
can be parsed.
Similarly, previous research also used
the~\mbox{(\topk{k})}~accuracy~\cite{Watson2020,Dinella2022,Tufano2022}
or the \bleu score~\cite{Tufano2022} to evaluate model performance.

\begin{figure*}[t]
  \centering
  \includegraphics[width=0.8\textwidth]{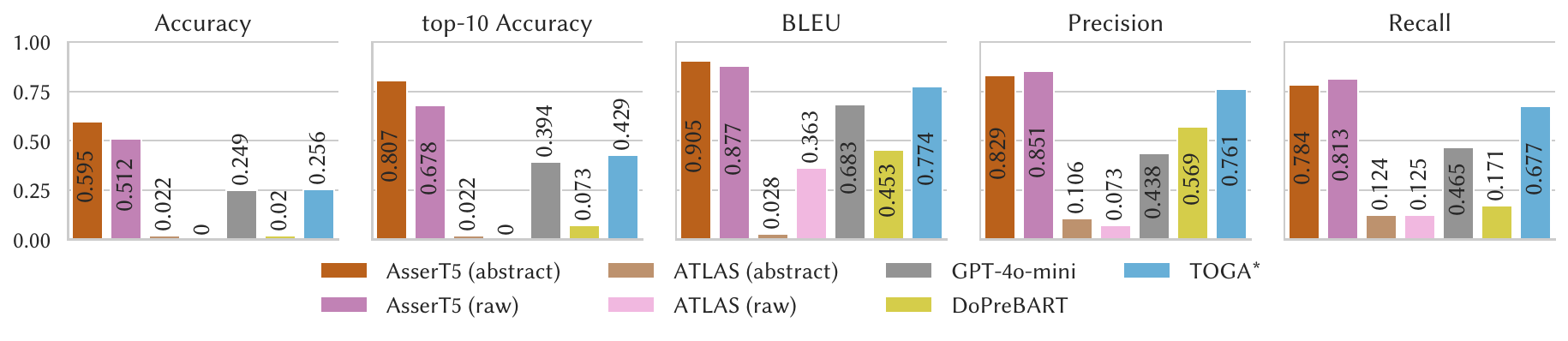}
  \caption{%
    Comparison of the individual models.
    \bleu score and accuracy compare the full assertion statement.
    The precision and recall scores only consider the assertion \emph{type}, \ie the \properNoun{JUnit} assertion method~(see \cref{tab:considered-assertion-types}).
  }\label{fig:model-comparison}
\end{figure*}

\subsubsection{Threats to Validity}

A threat to internal validity arises from the reconstruction of the
models we compare against. While we tried to implement the models as
close to their original as possible following their description, we
cannot guarantee that our implementations follow the original ones
exactly.
The training process of \chatgpt~\cite{Brown2020} highlights another
threat to internal validity. Since the model was trained on an
undisclosed large corpus of openly available data and our
\methodstotest-based dataset used for evaluation uses code mined from
open-source repositories from \github, we cannot ensure that the
training dataset of \chatgpt and our evaluation dataset are fully
distinct.
A threat to external validity may arise from the training process. We
relied mostly on the pre-determined hyperparameters present in the
replication packages for the comparison models and did not perform
further extensive tuning. There may be other parameter configurations
which improve model performance.
However, we expect the impact of different parameters to be limited,
since we either use the hyperparameters available in the replication
package~(\atlas) or reuse existing model architectures for
fine-tuning~(\bert, \bart) rather than training from scratch.

\subsubsection{Results}

\Cref{fig:model-comparison} shows the proportion of assertions that
the models predicted accurately. Both \assertfive models performed
better than the other models with the abstract variant performing best
by predicting the most assertions correctly with \perc{59.5}, followed
by the raw variant with \perc{51.2}.
The next-best model in our comparison is \togaast, with \perc{25.6} of
correctly predicted test assertions followed closely by
\chatgpt~(\perc{24.9}).

The accuracy of \atlas was only up to \perc{2.2}. We therefore cannot
confirm the results of~{Watson et
al.}~\cite{Watson2020}~(raw:~\perc{17.7}, abstract:~\perc{31.4}) with
our \atlas reimplementation. The results could be influenced by
training the models on a different dataset, using different
hyperparameters, or by a divergence in our reimplementation compared
to the original.
In our case \atlas only recognised a few different predictions. For
example, the raw model could only make twelve different predictions,
and \exnum{26315} predictions were always
\texttt{assertEquals(UNK, UNK.UNK())}, %
thus failing to find suitable replacement tokens.
Nevertheless, \assertfive still considerably outperforms the original
\atlas implementation and results~\cite{Watson2020}, and both the
\atlas with information retrieval extensions by Yu et
al.~\cite{Yu2022}~(accuracy: \perc{46.54}, \bleu: \perc{78.86}) and by
Sun et al.~\cite{Sun2023}~(accuracy: \perc{53.46}, \bleu:
\perc{80.77}), albeit on a different dataset compared to our
experiment.

When integrating an assertion suggestion tool, \eg,~as part of an \ide
plugin, it should show multiple alternative suggestions to the user.
The \topk{10} accuracy measures how often the actual assertion would
be part of the set of 10 suggestions, even in cases where the model
failed to predict the exact assertion as its first choice.
The score increase by \perc{67.6} compared to \topk{1} for \togaast
shows that this model generates a diverse set of assertions as part of
the \topk{10} suggestions. However, only in \perc{42.9} of the cases one
of them matches the original assertion.
The suggestions by the abstract \assertfive model may not be as
diverse~(\perc{35.6} score increase) but it generates more precise
assertions since one of the suggestions matches the original
\perc{80.7} of the time.
This may be more useful to a user in practice since they can
frequently select one of the suggestions without having to make
further changes to it.

The accuracy only considers the number of predictions that are fully
identical to the original. In practice, however, there are often
multiple alternative equivalent variants of an assertion like for
example \verb|assertEquals(0, list.size())| and
\verb|assertTrue(list.isEmpty())|. This could result in a model being
rated worse despite good quality performance in practice.
Such alternative assertions may explain why \chatgpt does not
outperform the other models even if it is much larger. Since the
training data of \chatgpt is not specific for assertion generation but
contains a broad spectrum of text data, the model may frequently
generate other similar assertions that do not entirely match the
expected one.

With the \bleu score such token-level similarities can be
measured~\cite{Papineni2002}.
As shown in \cref{fig:model-comparison}, \assertfive performed
similarly well in the raw and abstract variants with \bleu scores of
\perc{87.7} and \perc{90.5}, respectively.
Again, \togaast is the next-best model~(\perc{77.4}) followed by
\chatgpt~(\perc{68.3}).
\doprebart~(\perc{45.3}) and \atlas~(raw:~\perc{36.3},
abstract:~\perc{2.8}) obtained the lowest scores.

Focussing not on full assertion statements but only the assertion
types predicted by the model gives additional insights.
\Cref{fig:model-comparison} illustrates the precision and recall of
the assertion type classifications. The raw variant of \assertfive was
able to predict the assertion types with better
precision~(\perc{85.1}) and recall~(\perc{81.3}) than the abstract
variant~(precision: \perc{82.9}, recall: \perc{78.4}).
\togaast remained the third-strongest performer.

We found that the most frequent assertions in the dataset are
\texttt{assertEquals}~(\perc{59.34} of all assertions) and
\texttt{assertTrue}~(\perc{17.84}). To be useful, the models should
therefore be able to accurately generate such assertions.
This is represented by the prediction accuracy under the condition
that the assertion type was predicted correctly. In case of
\assertfive for \texttt{assertEquals}, this conditional
accuracy~(raw:~\perc{49.9}, abstract:~\perc{61.6}) is similar to the
overall accuracy. For \texttt{assertTrue} the conditional accuracy of
\assertfive even is substantially higher than the overall one for both
the raw~(\perc{66.7}) and the abstract~(\perc{76.7}) model.
This is in clear contrast to \doprebart where the model only predicts
the remainder of the assertion correctly in \perc{1.3} of cases when
it had suggested the assertion type \texttt{assertEquals} correctly
and therefore results in a low overall accuracy of \perc{2.0} even if
it can predict \perc{47.2} of \texttt{assertTrue} accurately.

\summary{2}{\assertfive outperforms the comparison models. Whereas the
abstract model predicts the assertions accurately, the raw model
predicts the assertion types more precisely.}

\subsection{RQ3: Bug-Detection Capability of Generated Assertions in
Developer-Written Tests}\label{sec:rq-defects4j}

To consider the actual practical applicability of the \assertfive
assertions in real-world projects, we use the bug database
\defectsforj~\cite{Just2014} to evaluate the bug-detecting
capabilities of both the raw and abstract \assertfive models by
combining developer-written tests with generated assertions.

\subsubsection{Experimental Setup}

\defectsforj~\cite{Just2014} is a database and extensible framework
that contains bugs that have occurred in real projects. The goal of
\defectsforj is to provide the software testing research community a
benchmark to compare new approaches reasonably.
We use six projects~(Chart, Closure, Lang, Math, Mockito, and Time)
from version 2.0.1 of the framework, containing a total of
\dfjTotalBugs non-deprecated bugs with \dfjTotalTests
developer-written bug-revealing test cases that fail with an assertion
failed error.
We remove test cases where inconsistencies~(\eg,~rare Unicode escapes
breaking regular expressions) arise during the experiment execution in
either the abstract or raw model variants.
Finally, we remove \dfjAssertionLocationInHelperTotal test cases where
the failing assertion is not located within the test itself but in
another helper method.
This results in a test corpus of
\dfjAssertionLocationInTestTotalBugs{} bugs with
\dfjAssertionLocationInTestTotalTests{} test cases.
Following previous work~\cite{Dinella2022,Tufano2022}, we generate an
assertion on the fixed version of the code. We therefore evaluate
whether \assertfive can be used to generate assertions for regression
tests.
An assertion generated by the model works as intended if it passes on
the fixed version of the code and fails on the buggy revision of the
production code.
To generate an assertion, we replace the originally failing assertion
with the placeholder~(see \cref{sec:data-preprocessing}) and pass the
test and focal method pair to \assertfive to generate a suitable
replacement.
The generated assertion is then placed back into the test case which
is executed against the buggy and fixed version of the production code
to check whether the assertion only fails on the buggy version.
By filtering the tests for ones that fail with an assertion error and
replacing the failing assertion, we ensure that the newly generated
assertion is the cause for the test failure rather than it being
caused by the test prefix.

Since the heuristics of \methodstotest~(see \cref{sec:methods2test})
were not sufficient to determine the focal method in all cases, we
extended them using additional information from the code difference
before and after the fix available in \defectsforj.
Candidates for the focal class are determined by matching the name to
the test class~(\ie, removing a \texttt{Test} affix from the name) and
by adding all classes that changed between the two code revisions.
Within those classes, we again check if a focal method can be found by
matching the name of the test case and a method in the production
code. To achieve this, we extract the subtokens from the names of the
test and focal methods and choose the method with the most common
subtokens as the focal method. If this fails, \ie,~there are no
intersecting subtokens, the last method call before the assertion is
chosen as focal method. In case this approach also proves
unsuccessful, we rely on the patch information to find changed
methods. In case the fix was located within a constructor, we treated
them like a proper focal method to still be able to provide some
context to the model.
For the one case where these automatic detection heuristics failed to
find a suitable method, we manually determined it by inspecting the
code.

\subsubsection{Threats to Validity}

A threat to the external validity arises from the \defectsforj
dataset.  Since it contains bugs from only six different open-source
projects which have been specifically sampled to be reproducible from
developer written test cases, it may not represent the structure of
test cases in other projects. However, the bugs in the dataset
represent real-world issues found in large open-source projects and
therefore are likely to be similar to bugs found in other projects.

Using the diff between bug-containing and fixed code for focal method
detection is an additional threat to external validity, since this is
only applicable in this specific dataset of \defectsforj. This option
is not feasible in real-world code scenarios. Given that predefined
heuristics for identifying the focal method from \methodstotest are
often inapplicable---such as for regression test names like
\texttt{testIssue1024}---we determined that using the diff provides a
valid alternative to at least find some suitable focal method
candidates.

We identify the exclusion of tests where the assertion is placed in a
helper method as threat to validity since this potentially enhances
the reported model performance.
Most of the test cases being removed for lacking a usable assertion
are part of the \emph{Closure} project. In
\dfjAssertionLocationInHelperClosure of \dfjTestsInProjectClosure
tests, the check for the expected result is placed in a separate helper
method, \eg,~\verb|checkCost("true", "1")| from \emph{Closure 28}.
For nearly all the excluded tests, the model would have produced an
uncompilable assertion due to using non-available variables or
methods, or generated a trivial assertion that passes in both the
buggy and failing code revisions. The bug detection capability
therefore remains nearly unchanged.
This threat highlights a research gap calling for alternative model
training methods: The strict pairing of only test and focal methods
does not always match the structure of real-world tests.
To provide \assertfive with the required context for the actual
assertion, it would be necessary to include all methods in the call
chain between the test method and the assertion as part of the input
to the model. Since the \assertfive model is trained on strict pairs
of test and focal methods, it has not been prepared during training
for the inclusion of such constructs.
Alternatively, by not including the helper methods in the input to
retrain the known structure, the test setup is obscured for the model
input. This holds especially in cases where the test case consists of
a single call to a helper method like frequently occurring in the
\emph{Closure} project, \eg,~only calling
\verb|testTypeCheck("JavaScript code")|.
Using \evosuite-generated tests like in previous
work~\cite{Dinella2022,Tufano2022} does not exhibit this limitation,
since the assertion never appears in helper methods in such tests.

\subsubsection{Results}

\begin{table}[t]
  \caption{%
    Developer-written test cases after replacing the original failing assertions with generated ones.}\label{tab:defects4j-assertion-results}
  \centering%
  \footnotesize%
  \begin{tabular}{lrr}
    \toprule
    Result & Abstract Model & Raw Model\\
    \midrule
    not compilable         & \dfjAbstractNotCompilable & \dfjRawNotCompilable\\
    fails on fixed         & \dfjAbstractFailsFixed    & \dfjRawFailsFixed\\
    fails only on buggy    & \dfjAbstractFailsBuggy    & \dfjRawFailsBuggy\\
    passes on both         & \dfjAbstractPassesBoth    & \dfjRawPassesBoth\\
    \bottomrule
  \end{tabular}
\end{table}

\cref{tab:defects4j-assertion-results} shows the distribution of the
\defectsforj results for both the abstract and the raw model variants.
A large share of the \dfjAssertionLocationInTestTotalTests{}
developer-written tests combined with the assertions generated by the
model are not compilable tests for both the
abstract~(\dfjAbstractNotCompilable) and the raw model
variant~(\dfjRawNotCompilable), resulting in
\dfjAbstractCompilableTestCases abstract and
\dfjRawCompilableTestCases raw compilable assertions. Of these,
\dfjAbstractFailsFixed abstract and \dfjRawFailsFixed raw assertions
fail on the fixed project variant. A test that has succeeded on the
fixed variant detects a bug only if it fails on the buggy variant of
\defectsforj. In total, there are \dfjAbstractFailsBuggy abstract and
\dfjRawFailsBuggy raw tests that could detect the bugs. The remaining
assertions~(abstract:~\dfjAbstractPassesBoth, raw:~\dfjRawPassesBoth)
pass both the fixed and the buggy variant. Consequently, they cannot
detect the specific bug identified by the \defectsforj dataset, but
may still be able to prevent future regressions.

In \defectsforj, every known bug has at least one bug-revealing test,
but multiple tests may reveal the same bug.  We define a bug as found
if it has at least one corresponding test case for which the generated
assertion fails only on the buggy version but passes on the fixed one.
Out of the \dfjAssertionLocationInTestTotalBugs{} bugs, the abstract
\assertfive model detected \dfjAbstractFoundBugs, while the raw model
identified \dfjRawFoundBugs bugs. In total, 14 were found by both
variants.

About one third of the tests no longer compiles after adding the
generated assertion.
Especially the raw model variants tends to sometimes generate
assertions requiring long String literals which are cut off at the
maximum output length of 64 tokens, resulting in missing closing
quotation marks or parentheses. Due to the shorter abstracted
constants, the abstract variant exhibits this problem only rarely.
Otherwise, most assertions are syntactically correct and calling
non-existing methods on objects is the common cause for the failing
compilation for both models.
However, constants for basic types like numbers or strings seem to be
handled correctly in comparisons or when appearing as parameters to
the assertion methods.

Considering the compiling assertions, only one third of them detect
the bugs~(abstract:~\perc{28.4}, raw:~\perc{29.9}). One reason for
this is that a large proportion of the predicted assertions already
fail in the fixed program variant. However, these often deviate only
slightly from the originally intended assertion, and may thus
nevertheless be helpful for developers.
For example, in the case of bug \emph{Chart 26}, the raw variant
predicted \verb|assertFalse(success)| and therefore the opposite of
the original assertion \verb|assertTrue(success)|.
Similarly, in the abstract variant for bug \emph{Math 91}, the
prediction only changed -1 appearing in the original assertion
\verb|assertEquals(-1, pi1.compareTo(pi2))| to 1.
The group of assertions that pass in both the buggy and fixed variant
also do not contribute to finding a bug. Trivial assertions that are
always fulfilled such as \verb|assertTrue(true)| often fall into this
category.
These examples illustrate the challenges involved in creating reliable
statements. On the one hand, even small deviations can cause the tests
to fail, even though they are consistent with the general intent,
while on the other hand, a too lenient assertion poses the risk of not
adequately capturing the intended functionality.

With \dfjRawFoundBugs out of \dfjAssertionLocationInTestTotalBugs bugs
being found, the bug-detection performance of \assertfive in our
experiment is worse compared to the evaluations of \toga~(finding 57
of 120)~\cite{Dinella2022}. However, the evaluations of \toga was
performed on \evosuite-generated rather than developer-written tests.
The structure of the automatically generated tests allows that the
heuristics as used during the \methodstotest training dataset
creation~(see \cref{sec:methods2test}) can be applied to determine the
focal method for the generated tests as well.
Since all tests where the focal method could not be determined were
removed from the \methodstotest data, both the training and evaluation
data is skewed towards tests following the required structure or
naming convention.
On the contrary, the developer-written tests used for our evaluation
do not have to conform to these requirements. We therefore expect that
our results are more representative of the performance when applying
the model in actual deployment scenarios.

Overall, while the model is able to successfully detect some bugs,
these results open up future research to improve upon various
limitations:
Firstly, some assertions are close to the original apart from swapped
comparisons or signs. In such cases \ide{}s could offer context-aware
actions that allow developers to quickly fix these instances.
Many of the remaining compilation errors could be fixed by models that
have a deeper understanding of the code semantics and can therefore
correctly apply more complex constructs.
Similarly, when designing such future more powerful models and
training them on more diverse inputs deviating from the strict
test/focal-method pairs~(\eg,~by additionally including
assertion-containing helper methods), the focus should not only be to
closely match the original assertion, but also evaluate whether the
deeper code-understanding helps to generate practically relevant
assertions.

\summary{3}{%
\assertfive detects \dfjRawFoundBugs of
\dfjAssertionLocationInTestTotalBugs bugs in our evaluation. While
this shows some promise towards the fault detection capabilities of
the model, the evaluation highlights practical limitations when
integrating the assertions into developer-written tests, thus
requiring future research.}

\section{Related Work}

The need for automatically generating test assertions first emerged in
the context of automated test generation. Since unit test generation
algorithms tend to focus on exploring sequences of calls, these
sequences need to be enhanced with assertions in order to help them
check for bugs other than unexpected exceptions or crashes. The Orstra
tool~\cite{Xie2006} introduced the idea to collect state information
while executing generated tests, and then instantiating assertion
templates based on the observed behaviour. By construction, these
assertions will pass on the code for which they were generated, such
that their main application lies in regression testing. Most
state-of-the art unit test generators such as
\evosuite~\cite{Fraser2011a}, \properNoun{Randoop}~\cite{Pacheco2007},
or \properNoun{Pynguin}~\cite{9793730} follow this approach. Since the
number of assertions that can be instantiated can be very large,
resulting in overly sensitive and unreadable test cases, test
assertions are often minimised using mutation analysis~\cite{6019060}.
The application of these techniques, however, has been limited to
automatically generated tests, rather than developer-written tests
which we focus on with our approach.

More recent assertion generation approaches use deep learning
techniques. While {Mastropaolo et al.} also present an approach based
on a smaller pre-trained \tfive base model, their evaluation only
reports \topk{k} accuracy metrics~\cite{Mastropaolo2021}.
By considering assertions to behave like regular statements, the next
test statement generation of \properNoun{TeCo}~\cite{Nie2023} can
generate assertions outperforming \atlas and \toga according to
accuracy and \bleu scores, but does not provide insights about their
performance on real bugs.

Many \llm{}-based approaches omit the task-specific model fine-tuning
and instead apply the large model directly to the task.
\properNoun{CEDAR}~\cite{Nashid2023} uses the \properNoun{CodeX} \llm
to demonstrate that prompting an \llm with a few examples of similar
focal and test method pairs before querying it with the actual request
yields more accurate assertions than \atlas.
Evaluating a similar \llm{}-based approach on Python rather than Java
code, \properNoun{CLAP}~\cite{Wang2024} achieves better performance
with zero-shot-prompting~(\ie,~no examples as part of the prompt).
The \togll~\cite{Hossain2024} approach compares various fine-tuned
code \llm{}s for the task of assertion generation. The evaluation on
artificially generated code mutastions shows that the strength of the
assertions in combination with \evosuite-generated tests is
significantly higher when compared to \toga~\cite{Hossain2024}.
In contrast to this paper, none of these approaches evaluate the
influence of the focal method or other assertions in the test method,
nor do they investigate the performance of the model when adding the
model-generated assertions to developer-written bug-detecting tests.
They also do not revisit the token abstraction process originally
proposed for the \properNoun{seq2seq}-based \atlas
model~\cite{Watson2020}. While it may no longer be necessary for
Transformer-based \llm{}s to overcome out-of-vocabulary
situations~\cite{Karampatsis2020}, our results show that it can
nevertheless be successfully applied to such models during fine-tuning
to generate more accurate assertions.

Rather than only generating assertions, \llm{}s have also been applied
to generate whole test methods for various commonly used programming
languages like JavaScript~\cite{Schaefer2024}, Python~\cite{Ryan2024},
Java~\cite{Chen2024}, or even supporting multiple
languages~\cite{He2024a}.
Besides closeness of the predicted tests to the original, their
evaluations frequently focus on the influence of the choice of
\llm~\cite{Schaefer2024,Ryan2024}, or
coverage~\cite{Schaefer2024,Ryan2024,Chen2024} rather than bug
detection capabilities of the generated tests.
We specifically focus on generating assertions rather than whole tests
with \assertfive. Developers might already have specific testing
scenarios in mind and have the expertise about the code semantics to
set up the test state accordingly. The assertion generation model then
acts in a supporting role to suggest possible additional checks that
might otherwise have been missed.

\section{Conclusions}

When developers have created a test scenario for a unit test they may
need help to select appropriate assertions to validate the resulting
program behaviour. One approach proposed in the literature is to
predict likely assertions using deep learning methods. While prior
results were already promising, open questions remained regarding the
influence of identifier abstraction, the context to include, the
benefits of using large pre-trained models of code, and the
effectiveness of assertions predicted for developer written tests at
revealing faults. To answer these questions, we introduce \assertfive,
a new model based on the pre-trained \codetfive model, and empirically
study assertion generation. In our experiments the \assertfive model
clearly outperforms prior models, and benefits from token abstraction
as well as additional context in the form of the method under test or
pre-existing assertions.

Our study also revealed several limitations adherent to deep
learning-based assertion generation techniques. In particular, even
though standard machine learning metrics suggest the predicted
assertions are accurate, they often nevertheless result in
uncompilable test code, or assert incorrect behaviour. While we
assumed a regression testing scenario in which assertions failing on
the code for which they are generated are problematic, there may be
potential for future research on using predicted assertions to find
faults already in the code.

The results of our study also confirm the importance of the inclusion
of a focal method in the context. This is an aspect that distinguishes
assertion prediction from related techniques that aim to predict
entire test cases: When the aim is to generate new tests, for example
using an \llm, then the user specifies the target method to be tested.
When adding assertions to existing test code, the focal method needs
to be determined automatically. Our experiments suggest that basic
heuristics are insufficient in practice, reinforcing the need for
research on focal method detection~\cite{Parizi2014,White2022,He2024}.

We provide implementations, training data and checkpoints for the
models, the raw data obtained during inference, and the evaluation
scripts at \url{https://doi.org/10.5281/zenodo.14703162}.

\section*{Acknowledgements}

This work is supported by Bayerische Forschungsstiftung under grant AZ-1520-21 (\enquote{DeepCode}).

\bibliographystyle{IEEEtran}
\bibliography{literature}

\begin{thebibliography}{10}
\providecommand{\url}[1]{#1}
\csname url@samestyle\endcsname
\providecommand{\newblock}{\relax}
\providecommand{\bibinfo}[2]{#2}
\providecommand{\BIBentrySTDinterwordspacing}{\spaceskip=0pt\relax}
\providecommand{\BIBentryALTinterwordstretchfactor}{4}
\providecommand{\BIBentryALTinterwordspacing}{\spaceskip=\fontdimen2\font plus
\BIBentryALTinterwordstretchfactor\fontdimen3\font minus
  \fontdimen4\font\relax}
\providecommand{\BIBforeignlanguage}[2]{{%
\expandafter\ifx\csname l@#1\endcsname\relax
\typeout{** WARNING: IEEEtran.bst: No hyphenation pattern has been}%
\typeout{** loaded for the language `#1'. Using the pattern for}%
\typeout{** the default language instead.}%
\else
\language=\csname l@#1\endcsname
\fi
#2}}
\providecommand{\BIBdecl}{\relax}
\BIBdecl

\bibitem{Yang2006}
\BIBentryALTinterwordspacing
Q.~Yang, J.~J. Li, and D.~Weiss, ``A survey of coverage based testing tools,''
  in \emph{International Workshop on Automation of Software Test (AST)}.\hskip
  1em plus 0.5em minus 0.4em\relax New York, NY, USA: ACM, 2006, pp. 99--103.
  [Online]. Available: \url{https://doi.org/10.1145/1138929.1138949}
\BIBentrySTDinterwordspacing

\bibitem{Beck2000}
\BIBentryALTinterwordspacing
K.~Beck and E.~Gamma, ``Test-infected: Programmers love writing tests,'' in
  \emph{More Java Gems}, ser. SIGS Reference Library, D.~Deugo, Ed.\hskip 1em
  plus 0.5em minus 0.4em\relax Cambridge University Press, 2000, pp. 357--376.
  [Online]. Available: \url{https://doi.org/10.1017/cbo9780511550881.029}
\BIBentrySTDinterwordspacing

\bibitem{Fraser2011}
\BIBentryALTinterwordspacing
G.~Fraser and A.~Arcuri, ``Evolutionary generation of whole test suites,'' in
  \emph{International Conference on Quality Software}, 2011, pp. 31--40.
  [Online]. Available: \url{https://doi.org/10.1109/qsic.2011.19}
\BIBentrySTDinterwordspacing

\bibitem{Pacheco2007}
\BIBentryALTinterwordspacing
C.~Pacheco and M.~D. Ernst, ``Randoop: Feedback-directed random testing for
  java,'' in \emph{ACM SIGPLAN Conference on Object-Oriented Programming
  Systems and Applications Companion (OOPSLA)}.\hskip 1em plus 0.5em minus
  0.4em\relax New York, NY, USA: ACM, 2007, pp. 815--816. [Online]. Available:
  \url{https://doi.org/10.1145/1297846.1297902}
\BIBentrySTDinterwordspacing

\bibitem{Nie2023}
\BIBentryALTinterwordspacing
P.~Nie, R.~Banerjee, J.~J. Li, R.~J. Mooney, and M.~Gligoric, ``Learning deep
  semantics for test completion,'' in \emph{IEEE/ACM International Conference
  on Software Engineering (ICSE)}.\hskip 1em plus 0.5em minus 0.4em\relax IEEE,
  2023. [Online]. Available: \url{https://doi.org/10.1109/icse48619.2023.00178}
\BIBentrySTDinterwordspacing

\bibitem{Schaefer2024}
\BIBentryALTinterwordspacing
M.~Sch\"afer, S.~Nadi, A.~Eghbali, and F.~Tip, ``An empirical evaluation of
  using large language models for automated unit test generation,'' \emph{IEEE
  Transactions on Software Engineering}, vol.~50, no.~1, pp. 85--105, 2024.
  [Online]. Available: \url{https://doi.org/10.1109/tse.2023.3334955}
\BIBentrySTDinterwordspacing

\bibitem{He2024}
\BIBentryALTinterwordspacing
Y.~He, J.~Huang, H.~Yu, and T.~Xie, ``An empirical study on focal methods in
  deep-learning-based approaches for assertion generation,'' \emph{Proceedings
  of the ACM on Software Engineering}, vol.~1, no. FSE, pp. 1750--1771, 2024.
  [Online]. Available: \url{https://doi.org/10.1145/3660785}
\BIBentrySTDinterwordspacing

\bibitem{Watson2020}
\BIBentryALTinterwordspacing
C.~Watson, M.~Tufano, K.~Moran, G.~Bavota, and D.~Poshyvanyk, ``On learning
  meaningful assert statements for unit test cases,'' in \emph{ACM/IEEE
  International Conference on Software Engineering (ICSE)}.\hskip 1em plus
  0.5em minus 0.4em\relax ACM, 2020. [Online]. Available:
  \url{https://doi.org/10.1145/3377811.3380429}
\BIBentrySTDinterwordspacing

\bibitem{Dinella2022}
\BIBentryALTinterwordspacing
E.~Dinella, G.~Ryan, T.~Mytkowicz, and S.~K. Lahiri, ``{TOGA}: a neural method
  for test oracle generation,'' in \emph{IEEE/ACM International Conference on
  Software Engineering (ICSE)}.\hskip 1em plus 0.5em minus 0.4em\relax New
  York, NY, USA: ACM, 2022, pp. 2130--2141. [Online]. Available:
  \url{https://doi.org/10.1145/3510003.3510141}
\BIBentrySTDinterwordspacing

\bibitem{Tufano2022}
\BIBentryALTinterwordspacing
M.~Tufano, D.~Drain, A.~Svyatkovskiy, and N.~Sundaresan, ``Generating accurate
  assert statements for unit test cases using pretrained transformers,'' in
  \emph{ACM/IEEE International Conference on Automation of Software Test
  (AST)}.\hskip 1em plus 0.5em minus 0.4em\relax ACM, 2022. [Online].
  Available: \url{https://doi.org/10.1145/3524481.3527220}
\BIBentrySTDinterwordspacing

\bibitem{Wang2024}
\BIBentryALTinterwordspacing
H.~Wang, H.~Hu, C.~Chen, and B.~Turhan, ``Chat-like asserts prediction with the
  support of large language model,'' 2024. [Online]. Available:
  \url{https://doi.org/10.48550/arxiv.2407.21429}
\BIBentrySTDinterwordspacing

\bibitem{Mastropaolo2021}
\BIBentryALTinterwordspacing
A.~Mastropaolo, S.~Scalabrino, N.~Cooper, D.~Nader~Palacio, D.~Poshyvanyk,
  R.~Oliveto, and G.~Bavota, ``Studying the usage of text-to-text transfer
  transformer to support code-related tasks,'' in \emph{IEEE/ACM International
  Conference on Software Engineering (ICSE)}.\hskip 1em plus 0.5em minus
  0.4em\relax IEEE, 2021. [Online]. Available:
  \url{https://doi.org/10.1109/icse43902.2021.00041}
\BIBentrySTDinterwordspacing

\bibitem{Bower1989}
\BIBentryALTinterwordspacing
M.~McCloskey and N.~J. Cohen, ``Catastrophic interference in connectionist
  networks: The sequential learning problem,'' in \emph{Psychology of Learning
  and Motivation}, ser. Psychology of Learning and Motivation, G.~H. Bower,
  Ed.\hskip 1em plus 0.5em minus 0.4em\relax Elsevier, 1989, vol.~24, pp.
  109--165. [Online]. Available:
  \url{https://doi.org/10.1016/s0079-7421(08)60536-8}
\BIBentrySTDinterwordspacing

\bibitem{Britz2017}
\BIBentryALTinterwordspacing
D.~{Britz}, A.~{Goldie}, T.~{Luong}, and Q.~{Le}, ``{Massive Exploration of
  Neural Machine Translation Architectures},'' Mar. 2017. [Online]. Available:
  \url{https://doi.org/10.48550/arXiv.1703.03906}
\BIBentrySTDinterwordspacing

\bibitem{Vaswani2017}
\BIBentryALTinterwordspacing
A.~Vaswani, N.~Shazeer, N.~Parmar, J.~Uszkoreit, L.~Jones, A.~N. Gomez,
  L.~Kaiser, and I.~Polosukhin, ``Attention is all you need,'' 2017. [Online].
  Available: \url{https://doi.org/10.48550/arxiv.1706.03762}
\BIBentrySTDinterwordspacing

\bibitem{Lewis2020}
\BIBentryALTinterwordspacing
M.~Lewis, Y.~Liu, N.~Goyal, M.~Ghazvininejad, A.~Mohamed, O.~Levy, V.~Stoyanov,
  and L.~Zettlemoyer, ``{BART:} denoising sequence-to-sequence pre-training for
  natural language generation, translation, and comprehension,'' in
  \emph{Annual Meeting of the Association for Computational Linguistics}.\hskip
  1em plus 0.5em minus 0.4em\relax Association for Computational Linguistics,
  2020. [Online]. Available:
  \url{https://doi.org/10.18653/v1/2020.acl-main.703}
\BIBentrySTDinterwordspacing

\bibitem{Devlin2019}
\BIBentryALTinterwordspacing
J.~Devlin, M.-W. Chang, K.~Lee, and K.~Toutanova, ``Bert: Pre-training of deep
  bidirectional transformers for language understanding,'' in \emph{Conference
  of the North American Chapter of the Association for Computational
  Linguistics: Human Language Technologies}.\hskip 1em plus 0.5em minus
  0.4em\relax Association for Computational Linguistics, 2019. [Online].
  Available: \url{https://doi.org/10.18653/v1/n19-1423}
\BIBentrySTDinterwordspacing

\bibitem{Raffel2019}
\BIBentryALTinterwordspacing
C.~Raffel, N.~Shazeer, A.~Roberts, K.~Lee, S.~Narang, M.~Matena, Y.~Zhou,
  W.~Li, and P.~J. Liu, ``Exploring the limits of transfer learning with a
  unified text-to-text transformer,'' 2019. [Online]. Available:
  \url{https://doi.org/10.48550/arxiv.1910.10683}
\BIBentrySTDinterwordspacing

\bibitem{Wang2021}
\BIBentryALTinterwordspacing
Y.~Wang, W.~Wang, S.~Joty, and S.~C.~H. Hoi, ``{C}ode{T}5: Identifier-aware
  unified pre-trained encoder-decoder models for code understanding and
  generation,'' in \emph{Conference on Empirical Methods in Natural Language
  Processing (EMNLP)}, M.-F. Moens, X.~Huang, L.~Specia, and S.~W.-t. Yih,
  Eds.\hskip 1em plus 0.5em minus 0.4em\relax Online and Punta Cana, Dominican
  Republic: Association for Computational Linguistics, Nov. 2021, pp.
  8696--8708. [Online]. Available:
  \url{https://aclanthology.org/2021.emnlp-main.685}
\BIBentrySTDinterwordspacing

\bibitem{Le2022}
\BIBentryALTinterwordspacing
H.~Le, Y.~Wang, A.~D. Gotmare, S.~Savarese, and S.~C.~H. Hoi, ``Coderl:
  Mastering code generation through pretrained models and deep reinforcement
  learning,'' in \emph{Advances in Neural Information Processing Systems},
  S.~Koyejo, S.~Mohamed, A.~Agarwal, D.~Belgrave, K.~Cho, and A.~Oh, Eds.,
  vol.~35.\hskip 1em plus 0.5em minus 0.4em\relax Curran Associates, Inc.,
  2022, pp. 21\,314--21\,328. [Online]. Available:
  \url{https://proceedings.neurips.cc/paper_files/paper/2022/file/8636419dea1aa9fbd25fc4248e702da4-Paper-Conference.pdf}
\BIBentrySTDinterwordspacing

\bibitem{Tufano2022a}
\BIBentryALTinterwordspacing
M.~Tufano, S.~K. Deng, N.~Sundaresan, and A.~Svyatkovskiy, ``Methods2test: A
  dataset of focal methods mapped to test cases,'' in \emph{International
  Conference on Mining Software Repositories (MSR)}.\hskip 1em plus 0.5em minus
  0.4em\relax ACM, 2022. [Online]. Available:
  \url{https://doi.org/10.1145/3524842.3528009}
\BIBentrySTDinterwordspacing

\bibitem{Tufano2020}
\BIBentryALTinterwordspacing
M.~Tufano, D.~Drain, A.~Svyatkovskiy, S.~K. Deng, and N.~Sundaresan, ``Unit
  test case generation with transformers and focal context,'' 2020. [Online].
  Available: \url{https://doi.org/10.48550/arxiv.2009.05617}
\BIBentrySTDinterwordspacing

\bibitem{Tufano2019}
\BIBentryALTinterwordspacing
M.~Tufano, J.~Pantiuchina, C.~Watson, G.~Bavota, and D.~Poshyvanyk, ``On
  learning meaningful code changes via neural machine translation,'' in
  \emph{IEEE/ACM International Conference on Software Engineering
  (ICSE)}.\hskip 1em plus 0.5em minus 0.4em\relax IEEE, 2019. [Online].
  Available: \url{https://doi.org/10.1109/icse.2019.00021}
\BIBentrySTDinterwordspacing

\bibitem{Tufano2018}
\BIBentryALTinterwordspacing
M.~Tufano, C.~Watson, G.~Bavota, M.~Di~Penta, M.~White, and D.~Poshyvanyk, ``An
  empirical investigation into learning bug-fixing patches in the wild via
  neural machine translation,'' in \emph{ACM/IEEE International Conference on
  Automated Software Engineering (ASE)}.\hskip 1em plus 0.5em minus 0.4em\relax
  New York, NY, USA: ACM, 2018, pp. 832--837. [Online]. Available:
  \url{https://doi.org/10.1145/3238147.3240732}
\BIBentrySTDinterwordspacing

\bibitem{Tufano2019a}
\BIBentryALTinterwordspacing
------, ``Learning how to mutate source code from bug-fixes,'' in \emph{IEEE
  International Conference on Software Maintenance and Evolution
  (ICSME)}.\hskip 1em plus 0.5em minus 0.4em\relax IEEE, 2019. [Online].
  Available: \url{https://doi.org/10.1109/icsme.2019.00046}
\BIBentrySTDinterwordspacing

\bibitem{Loshchilov2017}
\BIBentryALTinterwordspacing
I.~Loshchilov and F.~Hutter, ``Decoupled weight decay regularization,'' 2017.
  [Online]. Available: \url{https://doi.org/10.48550/arxiv.1711.05101}
\BIBentrySTDinterwordspacing

\bibitem{Karampatsis2020}
\BIBentryALTinterwordspacing
R.-M. Karampatsis, H.~Babii, R.~Robbes, C.~Sutton, and A.~Janes, ``Big code !=
  big vocabulary: open-vocabulary models for source code,'' in \emph{IEEE/ACM
  International Conference on Software Engineering (ICSE)}.\hskip 1em plus
  0.5em minus 0.4em\relax ACM, Jun. 2020, pp. 1073--1085. [Online]. Available:
  \url{https://doi.org/10.1145/3377811.3380342}
\BIBentrySTDinterwordspacing

\bibitem{Husain2019}
\BIBentryALTinterwordspacing
H.~Husain, H.-H. Wu, T.~Gazit, M.~Allamanis, and M.~Brockschmidt,
  ``Codesearchnet challenge: Evaluating the state of semantic code search,''
  2019. [Online]. Available: \url{https://doi.org/10.48550/arxiv.1909.09436}
\BIBentrySTDinterwordspacing

\bibitem{Brown2020}
\BIBentryALTinterwordspacing
T.~B. Brown, B.~Mann, N.~Ryder, M.~Subbiah, J.~Kaplan, P.~Dhariwal,
  A.~Neelakantan, P.~Shyam, G.~Sastry, A.~Askell, S.~Agarwal, A.~Herbert-Voss,
  G.~Krueger, T.~Henighan, R.~Child, A.~Ramesh, D.~M. Ziegler, J.~Wu,
  C.~Winter, C.~Hesse, M.~Chen, E.~Sigler, M.~Litwin, S.~Gray, B.~Chess,
  J.~Clark, C.~Berner, S.~McCandlish, A.~Radford, I.~Sutskever, and D.~Amodei,
  ``Language models are few-shot learners,'' 2020. [Online]. Available:
  \url{https://doi.org/10.48550/arxiv.2005.14165}
\BIBentrySTDinterwordspacing

\bibitem{Utkin2022}
\BIBentryALTinterwordspacing
I.~Utkin, E.~Spirin, E.~Bogomolov, and T.~Bryksin, ``Evaluating the impact of
  source code parsers on ml4se models,'' 2022. [Online]. Available:
  \url{https://doi.org/10.48550/arxiv.2206.08713}
\BIBentrySTDinterwordspacing

\bibitem{Yu2022}
\BIBentryALTinterwordspacing
H.~Yu, Y.~Lou, K.~Sun, D.~Ran, T.~Xie, D.~Hao, Y.~Li, G.~Li, and Q.~Wang,
  ``Automated assertion generation via information retrieval and its
  integration with deep learning,'' in \emph{IEEE/ACM International Conference
  on Software Engineering (ICSE)}.\hskip 1em plus 0.5em minus 0.4em\relax New
  York, NY, USA: ACM, 2022, pp. 163--174. [Online]. Available:
  \url{https://doi.org/10.1145/3510003.3510149}
\BIBentrySTDinterwordspacing

\bibitem{Sun2023}
\BIBentryALTinterwordspacing
W.~Sun, H.~Li, M.~Yan, Y.~Lei, and H.~Zhang, ``Revisiting and improving
  retrieval-augmented deep assertion generation,'' in \emph{IEEE/ACM
  International Conference on Automated Software Engineering (ASE)}.\hskip 1em
  plus 0.5em minus 0.4em\relax IEEE, Sep. 2023, pp. 1123--1135. [Online].
  Available: \url{https://doi.org/10.1109/ase56229.2023.00090}
\BIBentrySTDinterwordspacing

\bibitem{Papineni2002}
\BIBentryALTinterwordspacing
K.~Papineni, S.~Roukos, T.~Ward, and W.-J. Zhu, ``Bleu: a method for automatic
  evaluation of machine translation,'' in \emph{Annual Meeting on Association
  for Computational Linguistics}.\hskip 1em plus 0.5em minus 0.4em\relax USA:
  Association for Computational Linguistics, 2002, pp. 311--318. [Online].
  Available: \url{https://doi.org/10.3115/1073083.1073135}
\BIBentrySTDinterwordspacing

\bibitem{Just2014}
\BIBentryALTinterwordspacing
R.~Just, D.~Jalali, and M.~D. Ernst, ``{Defects4J}: A database of existing
  faults to enable controlled testing studies for java programs,'' in
  \emph{International Symposium on Software Testing and Analysis
  (ISSTA)}.\hskip 1em plus 0.5em minus 0.4em\relax New York, NY, USA: ACM,
  2014, pp. 437--440. [Online]. Available:
  \url{https://doi.org/10.1145/2610384.2628055}
\BIBentrySTDinterwordspacing

\bibitem{Xie2006}
\BIBentryALTinterwordspacing
T.~Xie, ``Augmenting automatically generated unit-test suites with regression
  oracle checking,'' in \emph{ECOOP}, 2006, pp. 380--403. [Online]. Available:
  \url{https://doi.org/10.1007/11785477_23}
\BIBentrySTDinterwordspacing

\bibitem{Fraser2011a}
\BIBentryALTinterwordspacing
G.~Fraser and A.~Arcuri, ``{EvoSuite}: Automatic test suite generation for
  object-oriented software,'' in \emph{ACM SIGSOFT Symposium and the European
  Conference on Foundations of Software Engineering}.\hskip 1em plus 0.5em
  minus 0.4em\relax New York, NY, USA: ACM, 2011, pp. 416--419. [Online].
  Available: \url{https://doi.org/10.1145/2025113.2025179}
\BIBentrySTDinterwordspacing

\bibitem{9793730}
\BIBentryALTinterwordspacing
S.~Lukasczyk and G.~Fraser, ``Pynguin: Automated unit test generation for
  python,'' in \emph{IEEE/ACM International Conference on Software Engineering
  (ICSE)}.\hskip 1em plus 0.5em minus 0.4em\relax ACM, 2022, pp. 168--172.
  [Online]. Available: \url{https://doi.org/10.1145/3510454.3516829}
\BIBentrySTDinterwordspacing

\bibitem{6019060}
\BIBentryALTinterwordspacing
G.~Fraser and A.~Zeller, ``Mutation-driven generation of unit tests and
  oracles,'' \emph{IEEE Transactions on Software Engineering}, vol.~38, no.~2,
  pp. 278--292, 2012. [Online]. Available:
  \url{https://doi.org/10.1109/TSE.2011.93}
\BIBentrySTDinterwordspacing

\bibitem{Nashid2023}
\BIBentryALTinterwordspacing
N.~Nashid, M.~Sintaha, and A.~Mesbah, ``Retrieval-based prompt selection for
  code-related few-shot learning,'' in \emph{IEEE/ACM International Conference
  on Software Engineering (ICSE)}.\hskip 1em plus 0.5em minus 0.4em\relax IEEE,
  2023. [Online]. Available: \url{https://doi.org/10.1109/icse48619.2023.00205}
\BIBentrySTDinterwordspacing

\bibitem{Hossain2024}
\BIBentryALTinterwordspacing
S.~B. Hossain and M.~Dwyer, ``Togll: Correct and strong test oracle generation
  with llms,'' May 2024. [Online]. Available:
  \url{https://doi.org/10.48550/arxiv.2405.03786}
\BIBentrySTDinterwordspacing

\bibitem{Ryan2024}
\BIBentryALTinterwordspacing
G.~Ryan, S.~Jain, M.~Shang, S.~Wang, X.~Ma, M.~K. Ramanathan, and B.~Ray,
  ``Code-aware prompting: A study of coverage-guided test generation in
  regression setting using {LLM},'' \emph{Proceedings of the ACM on Software
  Engineering}, vol.~1, no. FSE, pp. 951--971, 2024. [Online]. Available:
  \url{https://doi.org/10.1145/3643769}
\BIBentrySTDinterwordspacing

\bibitem{Chen2024}
\BIBentryALTinterwordspacing
Y.~Chen, Z.~Hu, C.~Zhi, J.~Han, S.~Deng, and J.~Yin, ``Chatunitest: A framework
  for llm-based test generation,'' in \emph{International Conference on the
  Foundations of Software Engineering (FSE)}.\hskip 1em plus 0.5em minus
  0.4em\relax ACM, 2024. [Online]. Available:
  \url{https://doi.org/10.1145/3663529.3663801}
\BIBentrySTDinterwordspacing

\bibitem{He2024a}
\BIBentryALTinterwordspacing
Y.~He, J.~Huang, Y.~Rong, Y.~Guo, E.~Wang, and H.~Chen, ``Unitsyn: A
  large-scale dataset capable of enhancing the prowess of large language models
  for program testing,'' 2024. [Online]. Available:
  \url{https://doi.org/10.48550/arxiv.2402.03396}
\BIBentrySTDinterwordspacing

\bibitem{Parizi2014}
\BIBentryALTinterwordspacing
R.~M. Parizi, S.~P. Lee, and M.~Dabbagh, ``Achievements and challenges in
  state-of-the-art software traceability between test and code artifacts,''
  \emph{IEEE Transactions on Reliability}, vol.~63, no.~4, pp. 913--926, 2014.
  [Online]. Available: \url{https://doi.org/10.1109/tr.2014.2338254}
\BIBentrySTDinterwordspacing

\bibitem{White2022}
\BIBentryALTinterwordspacing
R.~White and J.~Krinke, ``Tctracer: Establishing test-to-code traceability
  links using dynamic and static techniques,'' \emph{Empirical Software
  Engineering}, vol.~27, no.~3, 2022. [Online]. Available:
  \url{https://doi.org/10.1007/s10664-021-10079-1}
\BIBentrySTDinterwordspacing

\end{thebibliography}

\balance%

\end{document}